\documentclass[aps,twocolumn,showpacs,preprintnumbers,prl]{revtex4-1}

\usepackage{graphicx}

\usepackage{dcolumn}

\usepackage{bm}

\usepackage{amsmath}

\usepackage{amssymb}

\usepackage{booktabs}

\usepackage{color}

\usepackage{epstopdf}

\usepackage{epsfig}

\newcommand{\bra}[1]{\langle #1|}

\newcommand{\ket}[1]{|#1\rangle}

\newcommand{\Ignore}[1]{}

\begin{document}

\title{Acoustic plasmons in extrinsic free-standing graphene}

\author{M. Pisarra,$^1$ A. Sindona,$^1$ P. Riccardi,$^1$ V. M. Silkin,$^{2,3,4}$ and J. M. Pitarke$^{5,6}$}

\affiliation{
$^1$Dipartimento di  Fisica, Universita della Calabria and INFN - Gruppo collegato di Cosenza,
Via P. Bucci cubo 30 C, 87036 Arcavacata di Rende (CS), Italy  \\
$^2$Donostia International Physics Center, Paseo Manuel de Lardizabal 4, E-20018 Donostia - San Sebastian, Basque Country, Spain \\
$^3$Departamento de F\'{\i}sica de Materiales, Facultad de Ciencias Qu\'{\i}micas, UPV/EHU, Apartado 1072, E-20080 Donostia - San Sebastian, Basque Country, Spain \\
$^4$IKERBASQUE, Basque Foundation for Science, 48011 Bilbo, Basque Country, Spain \\
$^5$CIC nanoGUNE, Tolosa Hiribidea 76, E-20018 Donostia - San Sebastian, Basque Country, Spain\\
$^6$Materia Kondentsatuaren Fisika Saila and Centro Fisica Materiales CSIC-UPV/EHU, 644 Posta Kutxatila, E-48080 Bilbo, Basque Country, Spain
}

\date{18 April 2013}

\begin{abstract}

An acoustic plasmon is predicted to occur, in addition to the conventional two-dimensional~(2D) plasmon, as the collective motion of a system of two types of electronic carriers coexisting in the very same 2D band of {\it extrinsic} (doped or gated) graphene. The origin of this novel mode resides in the strong anisotropy that is present in the graphene band structure near the Dirac point. This anisotropy allows for the coexistence of carriers moving with two distinct Fermi velocities along the $\Gamma$K direction, which leads to two modes of collective oscillation: one mode in which the two types of electrons oscillate in phase with one another [this is the conventional 2D graphene plasmon, which at long wavelengths ($q\to 0$) has the same dispersion, $q^{1/2}$, as the conventional 2D plasmon of a 2D free electron gas], and the other mode found here corresponding to a low-frequency acoustic oscillation [whose energy exhibits at long wavelengths a linear dependence on the 2D wavenumber $q$] in which the two types of electrons oscillate out of phase. If this prediction is confirmed experimentally, it will represent the first realization of acoustic plasmons originated in the collective motion of a system of two types of carriers coexisting within the very same band.

\end{abstract}


\maketitle

Over the recent years, the interest in graphene has impressively grown
in both fundamental research and technological applications~\cite{Graphene_review1}.
This is due to the fact that graphene exhibits a good number of interesting properties,
related mainly to its {\it novel} electronic structure near the Fermi level represented by
the so-called Dirac cone. A major issue is represented in this case by the variation of the charge
carrier density, which is caused by several conditions including,
for example, the shape and defects of graphene flakes, charge transfer
processes with the supporting material~\cite{Doping_graph_metals},
chemical doping~\cite{Graph_chem_dop}, and the application of gating
potentials~\cite{Gate_graphene}. The appearance of a two-dimensional~(2D) sheet plasmon in graphene
adsorbed on a variety of supporting materials has been observed
in several experiments~\cite{Plasmon_Gra_SiC2,Plasmon_Gra_SiC,Plasmon_Gra_SiC3},
where the monolayer graphene happens to be doped by charge transfer to or from the substrate;
on the theoretical side, tight-binding calculations~\cite{wustnjp06,hwdaprb07,EPL_graphene} and
{\it ab-initio} calculations~\cite{yathprl11,gayussc11,denoprb13} have been able to reproduce a 2D sheet plasmon in extrinsic
(doped or gated) free-standing graphene.

In this letter, we present an {\it ab-initio} description of the energy-loss spectrum of both intrinsic (undoped and ungated) and extrinsic free-standing monolayer graphene.
Starting with pristine (intrinsic) graphene, we include the effect of electron
injection by simply up shifting the Fermi level from the Dirac point,
that is by working under the assumption that the graphene band structure is unaffected by doping.
We find that the strong anisotropy that is present in the graphene band structure near the Dirac
point allows for the coexistence of a majority of electrons moving with two different velocities along the $\Gamma$K direction, thus leading to a remarkable realization of the old idea~\cite{pines} that low-energy acoustic plasmons (whose energy exhibits a linear dependence on the wavenumber) should exist in the collective motion of a system of two types of electronic carriers. Our energy-loss calculations [which we carry out in the Random-Phase Approximation~(RPA)] clearly show the existence of a low-frequency acoustic oscillation (in which the two types of electrons oscillate out of phase), in addition to the conventional 2D graphene collective mode described in Refs.~\cite{wustnjp06,hwdaprb07,EPL_graphene}
(in which the two types of electrons oscillate in phase with one another).

We start with the following expression for the in-plane RPA complex dielectric matrix of a many-electron system consisting of
periodically repeated (and well separated) graphene 2D sheets (atomic units are used throughout, unless stated otherwise):
\begin{equation}
\label{eps_zz}
\epsilon_{{\bf g},{\bf g}'}({\bf q},\omega) = \delta_{{\bf g},{\bf g}'} -
v_{{\bf g},{\bf g}'}({\bf q})\sum_{g_z,g_z'}\chi^0_{{\bf G},{\bf G}'}({\bf q},\omega).
\end{equation}
Here, ${\bf G}$ is a three-dimensional (3D) reciprocal-lattice vector: ${\bf G}=\{{\bf g},g_z\}$,
${\bf g}$ and ${\bf q}$ being an in-plane 2D reciprocal lattice vector and an in-plane 2D wavevector, respectively.
$v_{{\bf g},{\bf g}'}({\bf q})=2\pi\delta_{{\bf g},{\bf g}'}/|{\bf q}+{\bf g}|$
and $\chi^0_{{\bf G},{\bf G}'}({\bf q},\omega)$ represents the 3D Fourier transform of
the density-response function of non-interacting electrons:
\begin{multline}\label{Adler-Wiser}
\chi_{{\bf G} {\bf G}'}^0({\bf q},\omega)=\frac{2}{\Omega} \sum_{{\bf k}}^{\rm BZ} \sum_{v\,c} (f_{v {\bf k}} - f_{c {\bf k} + {\bf q} } ) \\
\times \frac{\rho_{vc{\bf k},{\bf q}}({\bf G}) \rho_{vc{\bf k},{\bf q}}^*({\bf G}') } {\omega + \varepsilon_{v{\bf k} }  -
\varepsilon_{c  {\bf k} + {\bf q} }  + {\rm i}\eta }.
\end{multline}
In eq.\eqref{Adler-Wiser}, $\Omega$ represents a normalization volume, ${\bf k}$ is an in-plane 2D wavevector in the first Brillouin Zone (BZ),
$f_{v {\bf k}}$ and $f_{c {\bf k}+{\bf q}}$ are occupation numbers corresponding to states in the valence ($v$) and conduction ($c$) energy bands, respectively,
and $\rho_{vc {\bf k},{\bf q}}$ is a shorthand for the matrix element
$\bra{v {\bf k}} e^{-{\rm i}({\bf q}+ {\bf G})\cdot {\bf r}} \ket{c {\bf k}+{\bf q}}$,
$\varepsilon_{vc{\bf k}}$ and $|vc{\bf k}\rangle$ being the eigenvalues and eigenvectors of a single-particle Hamiltonian, which we take to be the Kohn-Sham (KS) Hamiltonian of Density-Functional Theory (DFT).

The inelastic scattering cross section corresponding to a process in which (after the scattering of external electrons or electromagnetic waves) an electronic excitation of wavevector ${\bf q}+{\bf g}$ (${\bf q}$ being a wavevector in the BZ) and energy
$\omega$ is created at the
graphene 2D sheet is proportional to the energy-loss function
${\rm Im}\left[-\epsilon_{{\bf g},{\bf g}}^{-1}({\bf q},\omega)\right]$.
Collective excitations (plasmons) are dictated by zeros in the real part of the macroscopic dielectric function
\begin{equation}
\epsilon_M({\bf q}+{\bf g},\omega)=1/\epsilon^{-1}_{{\bf g},{\bf g}}({\bf q},\omega)
\label{macroscopic}
\end{equation}
in an energy region where the imaginary part is small.

Our \emph{ab-initio} scheme begins with the KS eigenvalues and eigenvectors, which we calculate in the local-density approximation~(LDA) by using the Perdew-Zunger parametrization~\cite{pz}
of the uniform-gas correlation energy. We use a plane-wave basis set~(with a cut-off energy of 25 Hartrees) and a norm-conserving pseudopotential of the Troullier-Martins
type~\cite{TroullierMartins}. Our system is made by periodically repeated 2D graphene sheets separated by a distance
of $\sim 20\,{\rm \AA}$.
The BZ integration is carried out by using an unshifted
$60\times60\times1$ Monkhorst-Pack grid~\cite{MonkhPack},
which results in a $3600$ ${\bf k}$-point sampling of the BZ.
From the converged electron density, we calculate the KS single-particle energies and orbitals
on a denser ${\bf k}$-point mesh~($720\times720\times1$), including up to $60$~bands.
These KS energies and orbitals are plugged into Eq.~\eqref{Adler-Wiser}, which we use to obtain
the $\chi^0$ matrix with up to $\sim 500\,{\bf G}$-vectors.
The in-plane RPA complex dielectric matrix is then computed from Eq.~(\ref{eps_zz}).
For the wavevectors and energies of interest here (below the $\pi$ plasmon at $\sim 5\,{\rm eV}$),
stable results were obtained by including in Eq.~(\ref{Adler-Wiser}) 51 reciprocal-lattice vectors of the form ${\bf G}=\{0,g_z\}$.

\begin{figure}[t]
\includegraphics[width=0.49\textwidth]{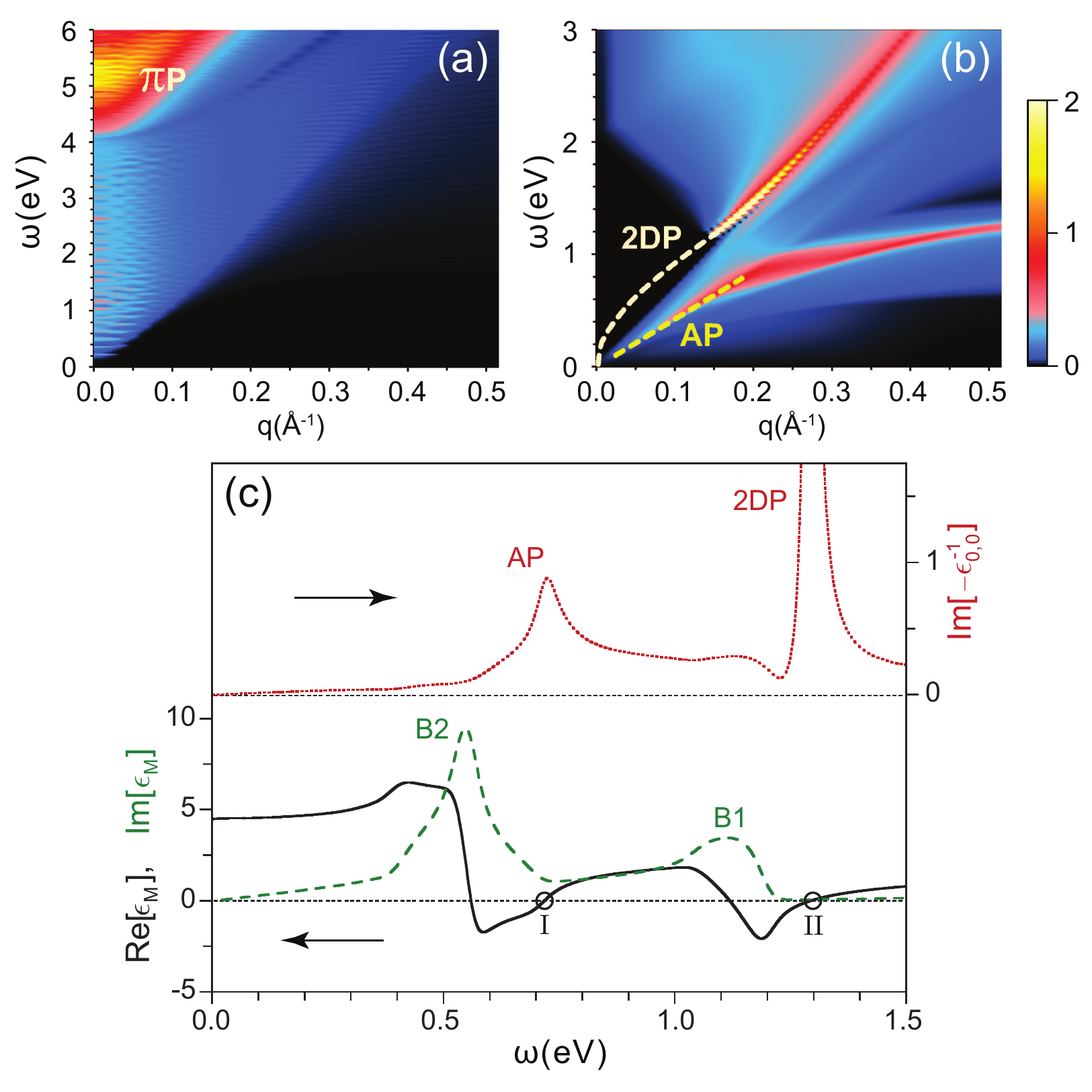}
\caption{2D plot of the energy-loss function of (a) {\it intrinsic} and (b) {\it extrinsic} graphene, {\it vs} the magnitude of the in-plane wavevector
${\bf q}$ along the $\Gamma$K direction (horizontal axis) and the energy $\omega$ (vertical axis). In the case of {\it extrinsic} graphene,
the Fermi level has been shifted $1\,{\rm eV}$ above the Dirac point.
(c) The energy-loss function ${\rm Im}\left[-\epsilon_{{\bf 0},{\bf 0}}^{-1}({\bf q},\omega)\right]$ (red line),
${\rm Re}\epsilon_M({\bf q},\omega)$ (black line), and ${\rm Im}\epsilon_M({\bf q},\omega)$ (green line) of {\it extrinsic}
graphene, {\it vs} the energy $\omega$ for a fixed value of the magnitude of ${\bf q}$ along the $\Gamma$K direction: $q=0.17\,{\rm \AA}^{-1}$.
\label{comp_doped_undoped_zoom}
}
\end{figure}

In the case of {\it intrinsic} graphene, the calculated energy-loss function presents three well-known distinct features. First of all, there is a broad peak-like structure starting at low values of $q$ and $\omega$ [see Fig.~\ref{comp_doped_undoped_zoom}(a)], which originates at interband $\pi\to\pi^*$ single-particle~(SP) excitations~\cite{wustnjp06,hwdaprb07} and was erroneously interpreted as a {\it cone plasmon} in Ref.~\cite{gayussc11}.
Second, there is the $\pi$ plasmon ($\pi$P) starting at $\sim 5\,{\rm eV}$ [see Fig.~\ref{comp_doped_undoped_zoom}(a)] (also present in graphite~\cite{taft}), which in the case of monolayer graphene is red-shifted and exhibits a linear dispersion~\cite{Olevano,himiepl09} distinct from the parabolic dispersion in graphite. Third, there is the broad high-energy graphene $\sigma-\pi$ plasmon peak starting at $\sim15$~eV [not visible in Fig. \ref{comp_doped_undoped_zoom}(a)], which corresponds to the graphite $\sigma-\pi$ plasmon at $\sim27$~eV~\cite{eels_novoselov}.

For {\it extrinsic} graphene, we adjust the occupation factors of Eq.~\eqref{Adler-Wiser} to account for a positive
Fermi-energy shift $\Delta E_F=1\,{\rm eV}$ relative to the Dirac point,
corresponding to a charge-carrier density of $1.15\times 10^{14}$~cm$^{-2}$~\cite{footnote_1}.
Figure~\ref{comp_doped_undoped_zoom} exhibits a comparison of the RPA energy-loss function that we have obtained along the $\Gamma$K direction for
{\it intrinsic} graphene~[Figure~\ref{comp_doped_undoped_zoom}(a)] and
{\it extrinsic} graphene~[Figure~\ref{comp_doped_undoped_zoom}(b)]. This doping affects neither the $\pi$ nor the $\sigma-\pi$ plasmon.
Important differences are visible, however, at low $q$ and $\omega$, where we can clearly identify
the opening of a gap in the SP excitation spectrum of {\it extrinsic} graphene. More importantly,
two collective modes (plasmons) are clearly visible in the case of {\it extrinsic} graphene (which are absent in {\it intrinsic} graphene):
(i) the conventional 2D graphene plasmon (2DP)~\cite{wustnjp06,hwdaprb07}, which within the gap (of the SP excitation spectrum) has no damping (and exhibits the same dispersion, $q^{1/2}$, as the conventional plasmon of a 2D electron gas~\cite{2D_system_review}) and outside the gap has finite linewidth, and (ii) a well-defined low-frequency mode (the {\it new} acoustic plasmon, AP), whose energy clearly exhibits at long wavelengths ($q\to 0$) a linear dependence on $q$.

In order to demonstrate that the energy-loss peaks that are visible in
Figure~\ref{comp_doped_undoped_zoom}(b) correspond to collective excitations, we have plotted in
Figure~\ref{comp_doped_undoped_zoom}(c) the energy-loss function (red line) for a given value of $q$ ($q=0.17$~\AA$^{-1}$), together with the real and the imaginary parts of the macroscopic dielectric function $\epsilon_M$ (black and green lines, respectively)
of Eq.~(\ref{macroscopic}). This figure clearly shows that ${\rm Re}\epsilon_M$ exhibits {\it two} distinct zeros (marked by the open circles I and II) in energy regions where ${\rm Im}\epsilon_M$ is small and the energy-loss function is, therefore, large. These two zeros ({\it each} of them being associated to the {\it two} maxima B1 and B2 in ${\rm Im}\epsilon_M$) represent a signature of well-defined
collective excitations: (i) The higher-energy plasmon (the conventional 2D graphene plasmon, 2DP) occurs at an energy
(just above the upper edge $v_Fq$ of the intraband SP excitation spectrum, $v_F$ being the graphene Fermi velocity) where only interband SP excitations are possible.
(ii) The low-energy plasmon (the {\it new} acoustic plasmon, AP) occurs at an energy that stays below $v_Fq$, so it is damped through intraband SP excitations; nonetheless, ${\rm Im}\,\epsilon_M$ is still considerably small at this energy, signaling that this low-energy mode represents a well-defined collective excitation as well.

\begin{figure}
\includegraphics[width=0.49\textwidth]{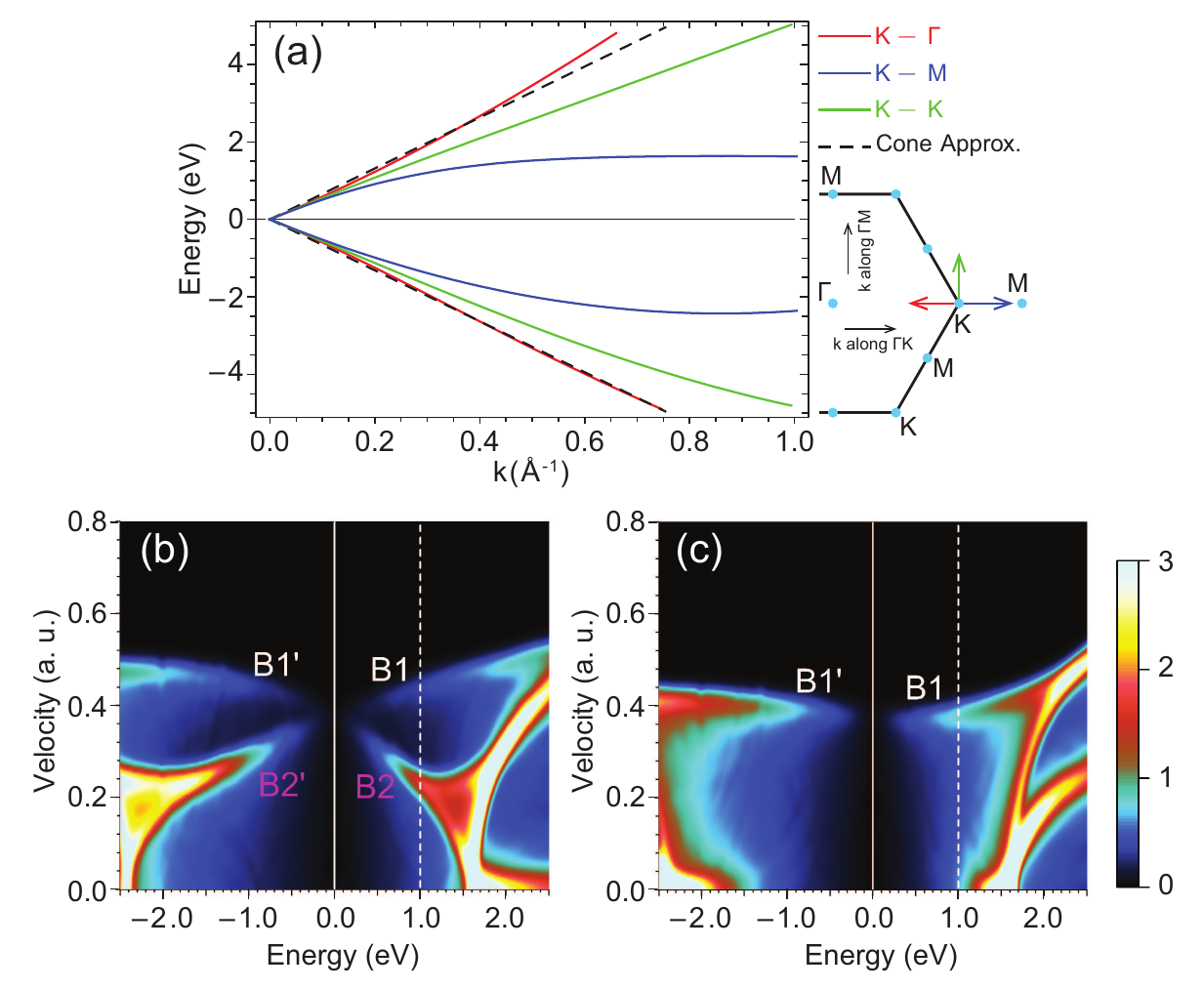}
\caption{(a) The graphene DFT band structure, as obtained along three high-symmetry paths all starting at the $K$-point:
The K$\Gamma$ and KM branches (red and blue lines, respectively) along the $\Gamma$K direction, and the
KK branch (green line) along the $\Gamma$M direction; the cone aproximation is represented by a black dashed line.
(b) and (c) Density of states, along $\Gamma$K and $\Gamma$M, {\it vs} the energy $\omega$ and the Bloch speed $v$.
The solid and dashed vertical lines represent the Fermi level of {\it intrinsic} graphene and {\it extrinsic}
graphene (with $\Delta E_F=1\,{\rm eV}$), respectively.
\label{bandsK_BZ}
}
\end{figure}

The existence of the low-energy acoustic plasmon could not possibly have been anticipated in the framework of simple
tight-binding-like investigations~\cite{wustnjp06,hwdaprb07}, simply because an oversimplified isotropic graphene band structure was considered in the vicinity of the K-point. A signature of such a mode has been detected recently~\cite{gayussc11,denoprb13}; but it was erroneously interpreted in Ref.~\cite{gayussc11} as a nonlinear mode along the nonlinear branch of the cone structure, and it was not discussed whatsoever in Ref.~\cite{denoprb13}.

With the aim of revealing the origin of the low-energy acoustic plasmon (the {\it new} plasmon), we show in Fig.~\ref{bandsK_BZ} the graphene
band structure [Figure~\ref{bandsK_BZ}(a)] and density of states [Figures~\ref{bandsK_BZ}(b) and \ref{bandsK_BZ}(c)] along various high-symmetry directions around the Dirac point.
Figure~\ref{bandsK_BZ}(a) shows our graphene band structure, as obtained along three high-symmetry paths all
starting at the K-point: The K$\Gamma$ and KM branches (red and blue lines, respectively) along the $\Gamma$K direction, and the
KK branch (green line) along the $\Gamma$M direction, together with the cone approximation (black dashed line).

The strong band-structure anisotropy that is visible in Fig.~\ref{bandsK_BZ}(a) implies the very unique behavior of the
density of states shown in Figs.~\ref{bandsK_BZ}(b) and \ref{bandsK_BZ}(c). While along the $\Gamma$M direction
[see Fig.~\ref{bandsK_BZ}(c)] the density of states is peaked (at the energies of interest, i.e., below $\sim 1.5\,{\rm eV}$) around one single Fermi velocity $v_F\sim 1\times 10^{6}\,{\rm m}/{\rm s}$ (peak B1 above the Dirac point and peak
B1' below), as occurs in a free-electron gas, the density of states along the $\Gamma$K direction is peaked at two distinct velocities
(peaks B1 and B2 above the Dirac point, and B1' and B2' below)
within the very same band. Since for a low wavevector along a given direction the number of allowed intraband transitions [dictated by
${\rm Im}\epsilon_M$] is known to be proportional to the density of states with group velocity along that direction~\cite{Pines_book}, intraband transitions along the $\Gamma$K direction [see also the maxima B1 and B2 in Figure~\ref{comp_doped_undoped_zoom}(c)] happen to be determined by the coexistence of carriers moving with two distinct Fermi velocities. This leads to two modes of collective oscillation:
(i) one mode (the conventional 2D plasmon, 2DP) in which the two types of electrons oscillate in phase with one another with
an energy that should be slightly larger than along the $\Gamma$M direction (where only one type of electrons participate and the 2DP dispersion -outside the gap- simply follows the upper intraband edge $v_Fq$~\cite{wustnjp06,hwdaprb07}),
and (ii) another mode (the {\it new} acoustic plasmon, AP) which corresponds to an {\it acoustic} oscillation of lower frequency in which the two types of electrons oscillate out of phase.

Hence, hereby we shed light on the observed deviation (along the $\Gamma$K direction) of the 2DP dispersion curve
towards energies that are (outside the gap) above the upper intraband edge $v_Fq$~\cite{Plasmon_Gra_SiC}. And hereby we predict the existence (along the $\Gamma$K direction) of a remarkable {\it acoustic} plasmon as the collective motion of a system of two types of electronic carriers coexisting in the very same 2D band of {\it extrinsic} graphene.

\begin{figure}[t]
\includegraphics[width=0.45\textwidth]{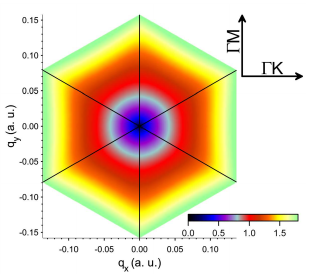}
\caption{The energy (in eV) of the conventional graphene 2D plasmon (2DP), {\it vs} the in-plane 2D wavevector {\bf q}. The Fermi level has been shifted $1\,{\rm eV}$ above the Dirac point, i.e., $\Delta E_F=1\,{\rm eV}$.
\label{2D_2DP_dispersion}
}
\end{figure}

\begin{figure}[t]
\includegraphics[width=0.45\textwidth]{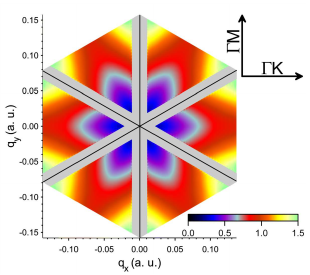}
\caption{As in Fig.~\ref{2D_2DP_dispersion}, but for the graphene acoustic plasmon (AP). The grey color shows regions
where the AP is not found to exist. For the ${\bf k}$ mesh and numerical broadening
used in our calculations, we have been able to trace the existence of the graphene AP down to $0.1\,{\rm eV}$.
\label{2D_AP_dispersion}
}
\end{figure}

The complete anisotropic plasmon dispersion of both plasmons (2DP and AP) is shown in
Figs.~\ref{2D_2DP_dispersion} and~\ref{2D_AP_dispersion}, respectively, where the plasmon energy is plotted {\it vs} the in-plane
2D wavevector {\bf q}. Figure~\ref{2D_2DP_dispersion} clearly shows that the conventional 2D plasmon is (i) isotropic at wavevectors below
$\sim0.1\,{\rm \AA}^{-1}$ ($0.05\,{\rm a.u.}$) where neither intraband nor iterband transitions are available and there is no damping, and
(ii) anisotropic at larger wavevectors (reflecting the 6-fold symmetry of the graphene BZ) with the plasmon energy being along the $\Gamma$K direction larger than along the
$\Gamma$M direction (as discussed above).

Figure~\ref{2D_AP_dispersion} shows that the {\it new} acoustic plasmon exhibits an extraordinary anisotropy. The
energy of the AP increases linearly with the magnitude of the wavevector, with a slope that is minimum along the $\Gamma$K direction and increases as one moves away from that direction until the AP completely disappears at wavevectors ${\bf q}$ along the $\Gamma$M direction (grey area).

In summary, we have demonstrated that as a consequence of the fact that two types of carriers in {\it extrinsic} graphene
(moving with two distinct Fermi velocities) coexist within the very same 2D band, (i) the conventional 2D plasmon
(corresponding to the two types of electrons oscillating in phase with one another)
disperses along the $\Gamma$K direction with an energy that is higher than along the $\Gamma$M direction, and
(ii) there is an additional {\it acoustic} plasmon (corresponding to the two types of electrons oscillating out of phase).
Low-energy acoustic plasmons are known to exist~\cite{sigaepl04,pinaprb04,dipon07,papaprl10,hellsing} at metal surfaces where a
quasi two-dimensional 2D surface-state band coexists with the underlying 3D continuum. Here we predict the existence of
a graphene acoustic plasmon, which if confirmed experimentally would be the first realization of acoustic plasmons originated
within the very same band and would represent, therefore, a truly remarkable feature having no analogue in solid-state physics.


\begin{thebibliography}{99}



\bibitem{Graphene_review1} A. H. Castro Neto, F. Guinea, N. M. R. Peres, K. S. Novoselov, and A. K. Geim, Rev. Mod. Phys. {\bf 81}, 109 (2009).



\bibitem{Doping_graph_metals} G. Giovannetti, P. A. Khomyakov, G. Brocks, V. M.  Karpan, J. van den Brink, and P. J. Kelly, Phys. Rev. Lett. {\bf 101}, 026803 (2008).



\bibitem{Graph_chem_dop} J. L. McChesney, A. Bostwick, T. Ohta, T. Seyller, K. Horn, J. Gonz\'alez,  and E. Rotenberg, Phys. Rev. Lett. {\bf 104}, 136803 (2010).



\bibitem{Gate_graphene} K. S. Novoselov, A. Geim, S. V.  Morozov,D. Jiang, Y. Zhang, S. V. Dubonos, I. V. Grigorieva, and A. A. Firsov, Science {\bf 306}, 666 (2004).



\bibitem{Plasmon_Gra_SiC2} Y. Liu, R. F. Willis, K. V. Emtsev, and T. Seyller, Phys. Rev. B {\bf 78}, 201403 (2008).



\bibitem{Plasmon_Gra_SiC}  C. Tegenkamp,H. Pfn\"ur, T. Langer, J. Baringhaus, and H. W. Schumacher, J. Phys: Condens Matter {\bf 23}, 012001 (2011).



\bibitem{Plasmon_Gra_SiC3} S. Y. Shin, C. G.  Hwang, S. J. Sung, N. D. Kim, H. S. Kim, and J. W. Chung, Phys. Rev. B {\bf 63}, 161403 (2011).



\bibitem{wustnjp06} B. Wunsch, T.  Stauber, F. Sols, and F. Guinea, New J. Phys. {\bf 8}, 318 (2006).



\bibitem{hwdaprb07} E. H. Hwang and S. Das Sarma, Phys. Rev. B {\bf 75}, 205418 (2007).



\bibitem{EPL_graphene} M. K. Kinyanjui, C. Kramberger, T. Pichler, J. C. Meyer, P. Wachsmuth, G. Benner, and U. Kaiser, Europhysics Letters {\bf 97}, 57005 (2012).



\bibitem{yathprl11} J. Yan, K. S. Thygesen, and K. W. Jacobsen, Phys. Rev. Lett. {\bf 106}, 146803 (2011).



\bibitem{gayussc11} Y. Gao and Z. Yuan, Solid State Commun. {\bf 151}, 1009 (2011).



\bibitem{denoprb13} V. Despoja, D. Novko, K. Dekani\'c, M. \v{S}unji\'{c}, and L. Maru\v{s}i\'{c}, Phys. Rev. B {\bf 87}, 075447 (2013).



\bibitem{pines} D. Pines, Can. J. Phys. {\bf 34}, 1379 (1956).



\bibitem{pz} J. P. Perdew and A. Zunger, Phys. Rev. B {\bf 23}, 5048 (1981).



\bibitem{TroullierMartins} N. Troullier and J. L. Martins, Phys. Rev. B {\bf 43}, 1993 (1991).



\bibitem{MonkhPack} H. J. Monkhorst and J. D. Pack, Phys. Rev. B {\bf 13}, 5188 (1976).



\bibitem{taft} E. A. Taft and H. R. Philipp, Phys. Rev. {\bf 138}, A197 (1965).



\bibitem{Olevano} C.  Kramberger, R. Hambach, C. Giorgetti, M. H. R\"ummeli, M. Knupfer, J. Fink, B. B\"uchner, L. Reining, E. Einarsson, S. Maruyama, {\it et al.}, Phys. Rev. Lett. {\bf 100}, 196803 (2008).



\bibitem{himiepl09} A. Hill, S. A. Mikhailov, and K. Ziegler, Europhys. Lett. {\bf 87}, 27005 (2009).



\bibitem{eels_novoselov} T. Eberlein, U. Bangert, R. R. Nair, R. Jones,  M. Gass, A. L. Bleloch, K. S. Novoselov, A. Geim, and P. R. Briddon, Phys. Rev. B {\bf 77}, 233406 (2008).



\bibitem{footnote_1} Charge-carrier densities achievable with gating potentials are typically of the order of

$10^{12}-10^{13}$~cm$^{-2}$~\cite{Gate_graphene}, so our analysis might be more

suitable for chemical doping~\cite{Graph_chem_dop}.



\bibitem{2D_system_review} T. Ando, A. B. Fowler, and F. Stern, Rev. Mod. Phys. {\bf 54}, 437 (1982).



\bibitem{Pines_book} D. Pines and P. Nozieres, {\it The Theory of Quantum Liquids} (Addison-Wesley, New York, 1989).



\bibitem{sigaepl04} V. M. Silkin, A. Garcia-Lekue, J. M. Pitarke, E. V. Chulkov, E. Zaremba, and P. M. Echenique, Europhys. Lett. {\bf 66}, 260 (2004).



\bibitem{pinaprb04} J. M. Pitarke, V. U. Nazarov, V. M. Silkin, E. V. Chulkov, E. Zaremba, and P. M. Echenique,  Phys. Rev. B {\bf 70}, 205403 (2004).



\bibitem{dipon07} B. Diaconescu, K. Pohl, L. Vattuone, L. Savio, P. Hofmann, V. M. Silkin, J.M. Pitarke, E. V. Chulkov,

P. M. Echenique, D. Farias, and M. Rocca, Nature (London) {\bf 448}, 57 (2007).



\bibitem{papaprl10} S. J. Park and R. E. Palmer, Phys. Rev. Lett. {\bf 105}, 016801 (2010).



\bibitem{hellsing} M. Jahn, M. M\"uller, M. Endlich, N. N\'e el, J. Kr\"oger, and B. Hellsing, Phys. Rev. B {\bf 86}, 085453 (2013).

\end{thebibliography}
\end{document}